\documentclass[aps,twocolumn,a4paper,nofootinbib,superscriptaddress]{revtex4}
\usepackage[dvips]{graphicx}
\usepackage{amsmath}
\usepackage{amssymb}

\begin{document}

\title{Virial theorem, boundary conditions, and pressure \\ for massless Dirac electrons}
\author{Alexey A. Sokolik}%
\email{asokolik@hse.ru}
\affiliation{Institute for Spectroscopy, Russian Academy of Sciences, 108840 Troitsk, Moscow, Russia}%
\affiliation{National Research University Higher School of Economics, 109028 Moscow, Russia}%
\author{Andrey D. Zabolotskiy}%
\email{zabolotskiy@vniia.ru}
\affiliation{Dukhov Automatics Research Institute (VNIIA), 127055 Moscow, Russia}%
\author{Yurii E. Lozovik}%
\email{lozovik@isan.troitsk.ru}
\affiliation{Institute for Spectroscopy, Russian Academy of Sciences, 108840 Troitsk, Moscow, Russia}%
\affiliation{National Research University Higher School of Economics, 109028 Moscow, Russia}%
\affiliation{Dukhov Automatics Research Institute (VNIIA), 127055 Moscow, Russia}%

\begin{abstract}
The virial and the Hellmann--Feynman theorems for massless Dirac electrons in a solid are derived and analyzed using
generalized continuity equations and scaling transformations. Boundary conditions imposed on the wave function in a
finite sample are shown to break the Hermiticity of the Hamiltonian resulting in additional terms in the theorems in
the forms of boundary integrals. The thermodynamic pressure of the electron gas is shown to be composed of the kinetic
pressure, which is related to the boundary integral in the virial theorem and arises due to electron reflections from
the boundary, and the anomalous pressure, which is specific for electrons in solids. Connections between the kinetic
pressure and the properties of the wave function on the boundary are drawn. The general theorems are illustrated by
examples of uniform electron gas, and electrons in rectangular and circular graphene samples. The analogous
consideration for ordinary massive electrons is presented for comparison.
\end{abstract}

\maketitle

\section{Introduction}
Discovery of graphene \cite{CastroNeto} and three-dimensional Dirac and Weyl semimetals \cite{Armitage}, where the electron
low-energy dynamics is described by the effective Dirac equation for massless particles, uncovered the new area of solid
state physics of Dirac materials \cite{Wehling}. Peculiar properties of these materials motivate researchers to
reconsider the conventional notions and models of quantum electron phenomena initially developed for massive electrons.
One of such notions is the virial theorem and related quantum theory of stress and pressure of electron gas
\cite{Marc,Nielsen,Maranganti,MartinPendas}.

The virial theorem for a system of interacting particles provides the relationship between average kinetic energy,
Coulomb interaction energy, and external pressure \cite{Marc}. Applications of the virial theorem in classical and
quantum statistical physics include estimation of the system properties, derivation of equations of state, checking
accuracy of quantum chemistry and density functional calculations etc. The quantum-mechanical virial theorem for a
system of ordinary massive electrons can be derived by using scaling transformations of an electron wave function
\cite{Fock,Lowdin} and spatial integration of the continuity equation for momentum density \cite{Ziesche,Godfrey}. The
pressure can be exerted on the system by Coulomb potentials of atomic nuclei \cite{Ziesche,Bader}, by a generic
external potential, or by impenetrable walls with the Dirichlet boundary conditions imposed on the electron wave
function. In the last case, the Hamiltonian of the system is Hermitian only in the subspace of wave functions
satisfying these conditions, and the scaling transformations drive the functions out of the Hermiticity domain
\cite{Abad}. The non-Hermiticity of the Hamiltonian in the presence of the scaling transformations results in emergence
of the additional term in the virial theorem, which is proportional to pressure and has a form of an integral of the
virial current density through the boundary \cite{Marc,Argyres,Bobrov,Srebrenik}. An alternative way to calculate
pressure as a response of the system energy to small volume changes relies on the Hellmann--Feynman theorem. The
non-Hermiticity of a Hamiltonian of enclosed system in the space of perturbed wave functions leads to emergence of the
boundary integral in the Hellmann--Feynman theorem as well \cite{Esteve,Argyres,Lowdin,Konstantinou,Srebrenik}. The
consistency between these two definitions of pressure (see Ref.~\cite{Marc}, p.~289) requires fulfilment of a specific
\emph{boundary relationship} for the wave function on the system boundary which relates its spatial derivative to its
derivative with respect to the boundary location \cite{Cottrell,Fernandez,Argyres}. Note that the virial theorem is a
particular case of more general stress theorem \cite{Nielsen,Maranganti,Godfrey,Bader}.

In the Dirac materials, the linear electron dispersion causes modification of the conventional virial theorem
\cite{Stokes,Sokolik}. Moreover, the momentum cutoff imposed at the bottom of the valence band in order to bound the system
energy from below leads to appearance of the additional term in the resulting generalized virial theorem
\cite{Sokolik}. However, the proper quantum-mechanical analysis of boundary contributions to the virial theorem for
massless Dirac electrons is still lacking. The electron wave function in these materials is multi-component and obeys
the boundary conditions which differ from the Dirichlet ones for massive electrons. For graphene, the infinite mass
\cite{Berry,McCann}, zigzag, and armchair \cite{Brey,Akhmerov,McCann} boundary conditions are used depending on the
lattice edge crystal structure. For three-dimensional Dirac and Weyl semimetals, various boundary conditions are
proposed \cite{Volkov,Hashimoto}. Other possible anomalies in scaling properties of a system of massless Dirac
electrons can also give rise to additional terms in the virial theorem \cite{Lin}.

In this paper, we derive the generalized virial and Hellmann--Feynman theorems for massless Dirac electrons, that
contain additional terms coming from the non-Hermiticity of the Hamiltonian in the presence of a system boundary and
from the momentum cutoff. Associating these terms with the pressure, we show that the thermodynamic pressure of Dirac
electrons is the sum of the \emph{kinetic pressure}, which is caused by reflections of electrons from the boundary, and
the \emph{anomalous pressure}, which is caused by redistribution of electron states during changes of the system area in
the presence of the momentum cutoff.

For the kinetic pressure, we show that the physically relevant boundary conditions imply the boundary relationship for
the wave function, which allows us to achieve consistency between definitions of the pressure based on the virial and
Hellmann--Feynman theorems. In addition to the total pressure, we calculate the local pressure on the boundary and
connect it with the kinetic part of the stress tensor. To illustrate the derived theorems, we consider several
particular examples: uniform electron gas in graphene, and rectangular and circular graphene flakes with the
appropriate boundary conditions. The calculations of the electron pressure in these examples are in agreement with the
generalized virial theorem.

The article is organized as follows. In Section~\ref{sec_massive} we consider the virial and Hellmann--Feynman theorems
for ordinary massive electrons, and derive the boundary relationship and the quantum-mechanical expressions for the
electron pressure. In Section~\ref{sec_Dirac} we provide the similar analysis for massless Dirac electrons and reveal
the important differences stemming from the momentum cutoff and from different form of boundary conditions. In
Section~\ref{sec_examples} we consider electrons in graphene samples of different geometries and demonstrate fulfilment
of the general theorems in these systems, and in Section~\ref{sec_conclusions} we summarize and discuss our results.
\ref{Appendix_A} and \ref{Appendix_B} are devoted to consideration of the scaling properties of uniform gases of
massive and massless electrons, which are closely connected with the virial theorems. In \ref{Appendix_C} we consider
two-band model of massive electrons and in \ref{Appendix_D} we calculate corrections to graphene electron gas
properties caused by electron dispersion nonlinearities.

\section{Massive electrons}\label{sec_massive}
\subsection{Virial theorem}
The virial and Hellmann--Feynman theorems for a bounded system of ordinary massive electrons \cite{Marc} can be derived
from the generalized continuity equations. We will assume that the system is surrounded by infinitely high potential
walls, so the wave function obeys the Dirichlet boundary conditions. Consider first a single particle with a stationary
state wave function $\psi(\mathbf{r})$ obeying the Schrodinger equation $H\psi=E\psi$ locally at the point
$\mathbf{r}$. Multiplying this equation by $\psi^*A$ from the left (where $A$ is some operator) and subtracting the
Hermitian conjugate equation $\psi^*H^+=E\psi^*$, which is multiplied by $A\psi$ from the right, we get
\begin{eqnarray}
-\psi^*[H,A]\psi+\psi^*(H-H^+)A\psi=0.\label{cont1}
\end{eqnarray}
Since the wave function $\psi$ can disobey the Dirichlet boundary condition after action of $A$, the Hamiltonian
bracketed between $\psi^*$ and $A\psi$ becomes, in general case, non-Hermitian. In coordinate representation this
non-Hermiticity is demonstrated only by the kinetic part $H_\mathrm{kin}=-\hbar^2\nabla^2/2m$ of $H$, and we can write
\begin{eqnarray}
\psi^*(H-H^+)A\psi=-i\hbar\:\mathrm{div}\,\mathbf{J}[A],\label{curr1}
\end{eqnarray}
where
\begin{eqnarray}
\mathbf{J}[A]=\frac\hbar{2mi}\psi^*(\nabla-\nabla^+)A\psi\label{curr2}
\end{eqnarray}
is the single-particle generalized current density of the quantity corresponding to the operator $A$, e.~g., the probability current if $A=1$ or the momentum current if $A=p_\mu$ \cite{Marc}. Hereafter we treat
$\nabla\equiv\overrightarrow\nabla$ and $\nabla^+\equiv\overleftarrow\nabla$ as the operators, which act on the
functions, respectively, to the right and to the left. Substituting (\ref{curr1})--(\ref{curr2}) in (\ref{cont1}),
integrating over the volume of the system $\Omega$ and applying the Gauss theorem, we get the formula
\begin{eqnarray}
-\frac{i}\hbar\langle\psi|[H,A]|\psi\rangle+\oint\limits_{\partial\Omega}d\mathbf{s}\cdot\mathbf{J}[A]=0,\label{cont2}
\end{eqnarray}
which equates total generation rate of the quantity $A$ in the system to the flux of this quantity out of the system in a stationary
state.

As $A$, we can take the virial (virial of momentum, to be more precise) operator
\begin{eqnarray}
G=\frac{\mathbf{r}\cdot\mathbf{p}+\mathbf{p}\cdot\mathbf{r}}2=-i\hbar\left(\mathbf{r}\cdot\nabla+\frac{D}2\right),
\label{G}
\end{eqnarray}
where $D$ is the space dimensionality. If the particle moves in the external potential $U_\mathrm{ext}(\mathbf{r})$,
then $H=H_\mathrm{kin}+U_\mathrm{ext}$ and $[H,G]=i\hbar(-2H_\mathrm{kin}+\mathbf{r}\cdot\nabla U_\mathrm{ext})$. With
$A=G$, Eq.~(\ref{cont2}) takes the form of the virial theorem with the boundary term:
\begin{eqnarray}
\langle\psi|-2H_\mathrm{kin}+\mathbf{r}\cdot\nabla
U_\mathrm{ext}|\psi\rangle+\oint\limits_{\partial\Omega}d\mathbf{s}\cdot\mathbf{J}[G]=0.\label{virial2}
\end{eqnarray}

\subsection{Generalized Hellmann--Feynman theorem}
Let us return to a single-particle Schrodinger equation $H|\psi\rangle=E|\psi\rangle$ written for the whole state
vector $|\psi\rangle$ and admit a small variation of the Hamiltonian $\delta H$ and/or boundary conditions, resulting
in a small variation $|\delta\psi\rangle$ of $|\psi\rangle$. Taking into account conservation of the wave function
normalization $\langle\delta\psi|\psi\rangle+\langle\psi|\delta\psi\rangle=0$, which implies
$\langle\delta\psi|H|\psi\rangle=-\langle\psi|H^+|\delta\psi\rangle$, we can write the variation of energy
$E=\langle\psi|H|\psi\rangle$ as
\begin{eqnarray}
\delta E=\langle\psi|\delta H|\psi\rangle+\langle\psi|H-H^+|\delta\psi\rangle.\label{delta_E}
\end{eqnarray}
Assuming the variation $\delta\lambda$ of some parameter as a physical origin of both $\delta H$ and
$|\delta\psi\rangle$, we get the Hellmann--Feynman theorem, generalized for the case of non-Hermitian Hamiltonian of a
bounded system:
\begin{eqnarray}
\frac{\partial E}{\partial\lambda}=\left\langle\psi\left|\frac{\partial
H}{\partial\lambda}\right|\psi\right\rangle+\left\langle\psi\left|\vphantom{\frac{\partial\psi}{\partial\lambda}}
H-H^+\right|\frac{\partial\psi}{\partial\lambda}\right\rangle.\label{HF1}
\end{eqnarray}
Using (\ref{curr1}) and the Gauss theorem, we can rewrite (\ref{HF1}) as
\begin{eqnarray}
\frac{\partial E}{\partial\lambda}=\left\langle\psi\left|\frac{\partial H}{\partial\lambda}\right|\psi\right\rangle-
i\hbar\oint\limits_{\partial\Omega}d\mathbf{s}\cdot\mathbf{J}\left[\frac\partial{\partial\lambda}\right].\label{HF2}
\end{eqnarray}
The boundary integrals in the virial (\ref{virial2}) and Hellmann--Feynman (\ref{HF2}) theorems for massive electrons
\cite{Marc} can be related to the pressure, as will be shown below.

\subsection{Thermodynamic pressure}
The thermodynamic pressure $\mathcal{P}=-\partial E/\partial\Omega$ is defined as a response of the system energy to
adiabatically slow volume change. The latter can be introduced as a uniform and isotropic dilation or contraction of
the system boundary points $\mathbf{r}\rightarrow\mathbf{r}(1+\delta R/R)$, where $R$ is a linear size of the system.
In this approach the Hamiltonian of the system does not change, and the wave function is affected only by the change of
boundary conditions. Taking $R$ as a slowly varying parameter in (\ref{HF2}), we obtain
\begin{eqnarray}
D\mathcal{P}\Omega=-R\frac{\partial E}{\partial R}=
i\hbar\oint\limits_{\partial\Omega}d\mathbf{s}\cdot\mathbf{J}\left[R\frac\partial{\partial R}\right].\label{p1}
\end{eqnarray}

An alternative way to calculate the pressure is to assume a finite-height confining potential
$U_\mathrm{b}(\mathbf{r})$ on the boundary added to the Hamiltonian, so we can discard the boundary integrals, because
the Hamiltonian becomes Hermitian due to vanishing of $\psi$ and $\nabla\psi$ at $|\mathbf{r}|\rightarrow\infty$.
Introducing the system size dependence $U_\mathrm{b}(\mathbf{r})=\tilde{U}_\mathrm{b}(\mathbf{r}/R)$ and applying the
ordinary virial (\ref{virial2}) and Hellmann--Feynman theorems, we get:
\begin{eqnarray}
\langle\psi|-2H_\mathrm{kin}+\mathbf{r}\cdot\nabla
U_\mathrm{ext}|\psi\rangle+D\mathcal{P}\Omega=0,\label{virial5}\\
D\mathcal{P}\Omega=\langle\psi|\mathbf{r}\cdot\nabla U_\mathrm{b}|\psi\rangle.\label{virial6}
\end{eqnarray}
In this paper, we do not use this method to define the system boundary, because it is inapplicable in the case of
massless Dirac electrons, which cannot be confined by a scalar potential \cite{CastroNeto} because their energy
spectrum is unbounded from below. Instead, for both kinds of electrons we define the boundary directly through the
boundary conditions for the wave function, giving rise to the boundary integral in (\ref{virial2}). Connection of this
integral with the physically measurable pressure will be drawn below.

\subsection{Boundary relationship}\label{sec_boundary}
\begin{figure}[t]
\begin{center}
\resizebox{0.5\columnwidth}{!}{\includegraphics{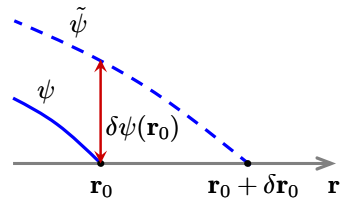}}
\end{center}
\caption{\label{Fig1} Wave functions near the boundary at initial $\mathbf{r}_0$ and perturbed
$\mathbf{r}_0+\delta\mathbf{r}_0$ boundary point locations, subject to the Dirichlet condition. The wave function
change $\delta\psi(\mathbf{r}_0)$ is shown by the arrows.}
\end{figure}

To calculate $\mathcal{P}$ using Eq.~(\ref{p1}), we need to know the derivative $\partial\psi/\partial R$ on the system
boundary. Suppose $\psi$ and $\tilde\psi$ are the wave functions of the same stationary state at, respectively, initial
and slightly perturbed boundaries (see Fig.~\ref{Fig1}). If any boundary point $\mathbf{r}_0$ moves outwards on a small
vector $\delta\mathbf{r}_0$, then the Dirichlet boundary conditions imply $\psi(\mathbf{r}_0)=0$ and
$\tilde\psi(\mathbf{r}_0+\delta\mathbf{r}_0)=0$, and the wave function change in a fixed point is
$\delta\psi(\mathbf{r}_0)\equiv\tilde\psi(\mathbf{r}_0)-\psi(\mathbf{r}_0)=\tilde\psi(\mathbf{r}_0)$. We can consider
$\psi(\mathbf{r})=\psi(\mathbf{r};\mathbf{r}_0)$ as a function of the vector $\mathbf{r}$ and the boundary position
$\mathbf{r_0}$, so $\tilde\psi(\mathbf{r})=\psi(\mathbf{r};\mathbf{r}_0+\delta\mathbf{r}_0)$, and
$\psi(\mathbf{r};\mathbf{r}_0)$ vanishes when its arguments coincide. Hence
$\delta\psi(\mathbf{r}_0)=\psi(\mathbf{r}_0;\mathbf{r}_0+\delta\mathbf{r}_0)=\delta\mathbf{r}_0\cdot
\nabla_{\mathbf{r}_0}\psi(\mathbf{r};\mathbf{r}_0)|_{\mathbf{r}=\mathbf{r}_0}+\mathcal{O}([\delta\mathbf{r}_0]^2)$. By
using the property $(\nabla_\mathbf{r}+\nabla_{\mathbf{r}_0})\psi(\mathbf{r};\mathbf{r}_0)=0$, we obtain the boundary
relationship
\begin{eqnarray}
\delta\psi(\mathbf{r}_0)=-\delta\mathbf{r}_0\cdot\nabla\psi(\mathbf{r}_0)+\mathcal{O}([\delta\mathbf{r}_0]^2).
\label{boundary1}
\end{eqnarray}
We can see from Fig.~\ref{Fig1} that $\delta\psi$ is indeed proportional to $\nabla\psi$ because the main cause of the
wave function change is just a motion of the boundary and the wave function as a whole, while the change of
$\nabla\psi$ provides only a second-order contribution.

Assuming a uniform and isotropic contraction or dilation of the system boundary
$\mathbf{r}_0\rightarrow\mathbf{r}_0(1+\delta R/R)$, we get from (\ref{boundary1}) another version of the boundary
relationship valid on the system boundary (see also \cite{Cottrell,Fernandez,Argyres}):
\begin{eqnarray}
R\frac{\partial\psi}{\partial R}=-\mathbf{r}\cdot\nabla\psi.\label{boundary2}
\end{eqnarray}
In \ref{Appendix_A} we show how this relationship can be obtained based on the scaling arguments. Using
(\ref{curr2}), (\ref{G}) (\ref{boundary2}) and the Dirichlet condition, we see that on the boundary
\begin{eqnarray}
\mathbf{J}[G]=i\hbar\,\mathbf{J}\left[\frac\partial{\partial R}\right].\label{boundary_currents}
\end{eqnarray}
This formula can be applied to unify (\ref{virial2}) and (\ref{p1}) as the virial theorem with the pressure
term:
\begin{eqnarray}
\langle\psi|-2H_\mathrm{kin}+\mathbf{r}\cdot\nabla
U_\mathrm{ext}|\psi\rangle+D\mathcal{P}\Omega=0,\label{virial7}\\
D\mathcal{P}\Omega=\oint\limits_{\partial\Omega}d\mathbf{s}\cdot\mathbf{J}[G]=\frac{\hbar^2}{2m}
\oint\limits_{\partial\Omega} d\mathbf{s}\cdot\mathbf{r}|\nabla\psi|^2\label{virial8}
\end{eqnarray}
(here we have used that Dirichlet boundary conditions imply that $\nabla\psi$ is directed parallel to the normal to the
boundary). It is similar to (\ref{virial5})--(\ref{virial6}), but formulated for a bounded system.

\subsection{Local pressure}\label{local_pressure_massive}
Now let us calculate from (\ref{delta_E}) the response of the system energy on arbitrary small perturbations
$\delta\mathbf{r}_0$ of the boundary points $\mathbf{r}_0$. Using (\ref{curr1})--(\ref{curr2}) with $A\psi=\delta\psi$
and applying the Gauss theorem, we get
\begin{eqnarray}
\delta E=-\frac{\hbar^2}{2m}\oint\limits_{\partial\Omega}d\mathbf{s}\cdot\psi^*\nabla^+\delta\psi.
\end{eqnarray}
The boundary relationship (\ref{boundary1}) allows us to rewrite it in the form
\begin{eqnarray}
\delta E=-\frac{\hbar^2}{2m}
\oint\limits_{\partial\Omega}ds_\nu\:(\delta\mathbf{r}_0)_\mu\psi^*\nabla_\nu^+\nabla_\mu\psi.\label{p2}
\end{eqnarray}
Introducing the kinetic stress tensor (or spatial part of the stress-energy tensor, or momentum flux density) by
applying (\ref{curr2}) to the momentum operator \cite{Maranganti},
\begin{eqnarray}
T_{\mu\nu}\equiv J_\nu[p_\mu]=\frac{\hbar^2}{2m}\psi^*\nabla_\mu(\nabla_\nu^+-\nabla_\nu)\psi,\label{stress_massive}
\end{eqnarray}
we recast (\ref{p2}) as
\begin{eqnarray}
\delta E=-\oint\limits_{\partial\Omega}ds_\nu\:(\delta\mathbf{r}_0)_\mu T_{\mu\nu}.\label{ploc0}
\end{eqnarray}
Associating $\delta E$ with the work done by external forces acting on the system, which are opposite to the
\emph{vector of local pressure} $\mathbf{P}$ of the system itself, we write
\begin{eqnarray}
\delta E=-\oint\limits_{\partial\Omega}ds\:\delta\mathbf{r}_0\cdot\mathbf{P},\label{ploc2}
\end{eqnarray}
thus, given the arbitrariness of $\delta\mathbf{r}_0$, the comparison of (\ref{ploc0}) and (\ref{ploc2}) results in
\begin{eqnarray}
P_\mu=T_{\mu\nu}n_\nu,\label{ploc1}
\end{eqnarray}
where $\mathbf{n}$ is the unit normal to the boundary and directed outside. The physical meaning of Eq.~(\ref{ploc1})
is that the system pressure $P_\mu$ exerted to the surroundings at some point is caused by particle collisions with the
boundary, which transfer momentum at the rate proportional to the normal component $\mathbf{J}[p_\mu]\cdot\mathbf{n}$
of the momentum flux at that point.

In the case of uniform dilation $\delta\mathbf{r}_0=(\delta R/R)\mathbf{r}_0$ of the system boundary we recover
(\ref{virial8}) with the relationship between thermodynamic and local pressures:
\begin{eqnarray}
D\mathcal{P}V=\oint\limits_{\partial\Omega}ds\:\mathbf{r}\cdot\mathbf{P}= \oint\limits_{\partial\Omega}ds_\nu\:r_\mu
T_{\mu\nu}.\label{ploc_integral}
\end{eqnarray}
The integrals here are independent of the choice of the origin because $\oint_{\partial\Omega}d\mathbf{s}\cdot\mathbf{P}=0$ for a
system being in mechanical equilibrium with its surroundings.

\subsection{Many-body system}\label{sec_manybody}
The many-body and thermal ensemble generalizations of all calculations presented above are rather straightforward.
Assume that the system state is characterized by an $N$-particle density matrix
\begin{eqnarray}
\rho_N=\sum_nw_n|\Psi_n\rangle\langle\Psi_n|,\label{dens}
\end{eqnarray}
where $|\Psi_n\rangle$ are the eigenstates, $H|\Psi_n\rangle=E_n|\Psi_n\rangle$, of the many-body Hamiltonian
$H=\sum_i[(-\hbar^2\nabla_i^2/2m)+U_\mathrm{ext}(\mathbf{r}_i)]+(1/2)\sum_{i\neq
j}V_\mathrm{int}(\mathbf{r}_i-\mathbf{r}_j)$ with the energies $E_n$, entering the ensemble with the probabilities
$w_n$. As specific examples, we can consider the many-body ground state $|\Psi_0\rangle$ at $T=0$, where
$w_n=\delta_{n0}$, or the thermal state, where $w_n\propto\exp(-E_n/T)$.

Introducing the one-body density matrix
\begin{eqnarray}
\rho_1(\mathbf{r},\mathbf{r}')=\sum_nw_n\int\prod_jd\mathbf{r}_jd\mathbf{r}_j' \sum_i\delta(\mathbf{r}-\mathbf{r}_i)
\nonumber\\ \times\delta(\mathbf{r}'-\mathbf{r}_i')\Psi_n(\mathbf{r}_1\ldots \mathbf{r}_N)\Psi_n^*(\mathbf{r}_1'\ldots
\mathbf{r}_N'),
\end{eqnarray}
we can define the many-body counterpart of the generalized current (\ref{curr2}):
\begin{eqnarray}
\mathbf{J}[A]=\frac\hbar{2mi}\left.(\nabla_\mathbf{r}-\nabla_{\mathbf{r}'})A_\mathbf{r}
\rho_1(\mathbf{r},\mathbf{r}')\right|_{\mathbf{r}'=\mathbf{r}}.\label{curr3}
\end{eqnarray}
Another distinction of the many-body system is the presence of the interparticle interaction $V_\mathrm{int}$, leading
to an additional term $\propto[H,V_\mathrm{int}]$ in the virial theorem. For the Coulomb interaction, we have
$[H,V_\mathrm{int}]=-i\hbar V_\mathrm{int}$, and the many-body virial theorem can be obtained by taking the linear
combination of the single-particle ones (\ref{virial2}) with the coefficients $w_n$:
\begin{eqnarray}
\langle-2H_\mathrm{kin}-V_\mathrm{int}+\mathbf{r}\cdot\nabla
U_\mathrm{ext}\rangle+\oint\limits_{\partial\Omega}d\mathbf{s}\cdot\mathbf{J}[G]=0\label{virial3}
\end{eqnarray}
(here $\langle A\rangle$ is defined as $\mathrm{Tr}\,[\rho A]$). The generalized Hellmann--Feynman theorem (\ref{HF1})
for a many-body system is
\begin{eqnarray}
\frac{\partial
E}{\partial\lambda}=\sum_n\left\langle\Psi_n\left|\frac{\partial(w_nH)}{\partial\lambda}\right|\Psi_n\right\rangle-
i\hbar\oint\limits_{\partial\Omega}d\mathbf{s}\cdot\mathbf{J}\left[\frac\partial{\partial\lambda}\right].\label{HF3}
\end{eqnarray}

The thermodynamics pressure $\mathcal{P}$ can be calculated using (\ref{HF3}) with $\lambda=R$. At $T=0$ we have
$\mathcal{P}=-\partial E_0/\partial\Omega$, but at $T>0$ we need to consider the free energy $F$ to define the pressure
$\mathcal{P}=-\partial F/\partial\Omega$. In both cases the derivatives of $w_n$ in (\ref{HF3}) do not appear in the
resulting formula for $\mathcal{P}$, and we obtain
\begin{eqnarray}
D\mathcal{P}\Omega=i\hbar\oint\limits_{\partial\Omega}d\mathbf{s}\cdot\mathbf{J}\left[R\frac\partial{\partial
R}\right],\label{p4}
\end{eqnarray}
which looks equivalent to (\ref{p1}), although with the many-body current operator (\ref{curr3}).

The Dirichlet boundary conditions imposed on the many-body wave function $\Psi_n(\mathbf{r}_1\ldots\mathbf{r}_N)$ imply
$\Psi_n=0$ when any of $\mathbf{r}_i$ is located on the boundary. Therefore the same boundary relationship
(\ref{boundary1}) is valid for $\Psi_n$ when $\mathbf{r}$ is replaced by any of its arguments $\mathbf{r}_i$, and
Eq.~(\ref{boundary_currents}) is valid for a many-body system as well. Combining (\ref{boundary_currents}),
(\ref{curr3}), (\ref{virial3}), and (\ref{p4}), we obtain the many-body counterpart of the single-particle virial
theorem (\ref{virial7})--(\ref{virial8}) with the pressure term:
\begin{eqnarray}
\langle-2H_\mathrm{kin}-V_\mathrm{int}+\mathbf{r}\cdot\nabla
U_\mathrm{ext}\rangle+D\mathcal{P}\Omega=0,\label{virial10}\\
D\mathcal{P}\Omega=\oint\limits_{\partial\Omega}d\mathbf{s}\cdot\mathbf{J}[G]=\frac{\hbar^2}{2m}
\oint\limits_{\partial\Omega} ds_\nu\:r_\nu\nabla_\mu\nabla_\mu'\rho_1(\mathbf{r},\mathbf{r}').
\end{eqnarray}
Derivation of
(\ref{virial10}) in the case of power-law $U_\mathrm{ext}$ by using scaling properties is shown in
\ref{Appendix_A}. The consideration (\ref{ploc0})--(\ref{ploc1}) of the local pressure can be also repeated
with the many-body kinetic stress tensor
\begin{eqnarray}
T_{\mu\nu}=\frac{\hbar^2}{2m}\psi^*\nabla_\mu\left.(\nabla_\nu'-\nabla_\nu)
\rho_1(\mathbf{r},\mathbf{r}')\right|_{\mathbf{r}'=\mathbf{r}}.
\end{eqnarray}
Alongside with the local pressure on the boundary (\ref{ploc1}), we can define the local kinetic pressure
\cite{MartinPendas,Ziesche,Godfrey}
\begin{eqnarray}
\mathcal{P}_\mathrm{bulk}(\mathbf{r})=\frac1D\left(T_{\mu\mu}-\sigma_{\mu\mu}^\mathrm{int}+r_\mu
f_\mu^\mathrm{ext}\right)\label{p_bulk}
\end{eqnarray}
in the bulk related to a trace of the total stress tensor consisting of the kinetic part $T$, interaction part
$\sigma^\mathrm{int}$ defined in an appropriate gauge \cite{MartinPendas,Nielsen,Godfrey} and the contribution of the
external body forces acting on electrons with the spatial density
$f_\mu^\mathrm{ext}=-\rho_1(\mathbf{r},\mathbf{r})\nabla_\mu U_\mathrm{ext}(\mathbf{r})$. According to the stress
theorem \cite{Nielsen,Maranganti,Godfrey,Bader}, the spatial average of (\ref{p_bulk}) should be equal to the
thermodynamic pressure:
\begin{eqnarray}
\mathcal{P}=\frac1\Omega\int d\mathbf{r}\:\mathcal{P}_\mathrm{bulk}(\mathbf{r}).\label{p_bulk_theorem}
\end{eqnarray}

\section{Massless Dirac electrons}\label{sec_Dirac}

\subsection{Generalized virial and Hellmann--Feynman theorems}
Now we will turn to massless Dirac electrons in a solid. As a specific example, we consider two-dimensional system of
electrons in graphene, but our general theorems should be applicable to any other Dirac materials. The massless
electrons in graphene \cite{CastroNeto} have the following distinctions from the massive ones, important for our
analysis: 1) their effective (single-particle) wave function
$\psi=(\psi_{A\mathbf{K}},\psi_{B\mathbf{K}},\psi_{B\mathbf{K}'},\psi_{A\mathbf{K}'})^T$ is a multi-component column
with the components corresponding to the sublattices $A,B$ and valleys $\mathbf{K},\mathbf{K}'$; 2) the kinetic part of
the Hamiltonian is $H_\mathrm{kin}=v_\mathrm{F}\boldsymbol\Sigma\cdot\mathbf{p}$, where
$\boldsymbol\Sigma=\mathrm{diag}(\boldsymbol\sigma,-\boldsymbol\sigma)$ is the $(4\times4)$ vector matrix with the
vectors composed of Pauli matrices on the diagonal; 3) the boundary conditions imposed on the wave function are not the
Dirichlet condition but have more diverse forms of the system of equations $M\psi=\psi$, with the $(4\times4)$ matrices
$M$ dependent on the edge types \cite{Brey,Akhmerov,Berry,McCann}; 4) to define the ground state and make the system
energy bounded from below, an appropriate momentum cutoff $|\mathbf{p}|<p_\mathrm{c}$ deep in the valence band should
be introduced; this approach allows us to approximate the tight-binding model of electrons in graphene, which have a
physically bounded valence band, by a simpler effective model.

Now we can repeat the calculations of Section~\ref{sec_massive} with taking into account that for massless Dirac
electrons the single-particle
\begin{eqnarray}
\mathbf{J}[A]=v_\mathrm{F}\psi^+\boldsymbol\Sigma A\psi\label{curr4}
\end{eqnarray}
and many-body
\begin{eqnarray}
\mathbf{J}[A]=v_\mathrm{F}\left.\mathrm{Tr}\,[\boldsymbol\Sigma
A_\mathbf{r}\rho_1(\mathbf{r},\mathbf{r}')]\right|_{\mathbf{r}'=\mathbf{r}}\label{curr5}
\end{eqnarray}
expressions for the generalized current are different from (\ref{curr2}), (\ref{curr3}) due to different form of the
non-Hermitian $H_\mathrm{kin}$. The single-particle density matrix $\rho_1$ in (\ref{curr5}) is assumed to be a
$(4\times4)$ matrix over sublattice and valley degrees of freedom. Taking also into account that $H_\mathrm{kin}$ is
linear in momentum and assuming Coulomb interaction in a many-body system, that imply
$[H,G]=i\hbar(-H_\mathrm{kin}-V_\mathrm{int}+\mathbf{r}\cdot\nabla U_\mathrm{ext})$, we get the counterpart of the
virial theorem (\ref{virial3}) for massless Dirac electrons:
\begin{eqnarray}
\langle-H_\mathrm{kin}-V_\mathrm{int}+\mathbf{r}\cdot\nabla
U_\mathrm{ext}\rangle+\oint\limits_{\partial\Omega}d\mathbf{s}\cdot\mathbf{J}[G]=0.\label{virial4}
\end{eqnarray}
Eq.~(\ref{virial4}) can be compared with the generalized virial theorem, obtained in \cite{Sokolik} by means of scaling
transformations of a many-body wave function with the imposed momentum cutoff:
\begin{eqnarray}
\langle-H_\mathrm{kin}-V_\mathrm{int}+\mathbf{r}\cdot\nabla
U_\mathrm{ext}\rangle+D\mathcal{P}\Omega+p_\mathrm{c}\frac{\partial E}{\partial p_\mathrm{c}}=0\label{virial9}
\end{eqnarray}
(see the alternative derivation in \ref{Appendix_B}). Here, as in the previous sections,
$\mathcal{P}=-\partial E/\partial\Omega$, and $E$ should be understood as the ground state energy $E_0$ at $T=0$ or as
the free energy $F$ at $T>0$. Comparing (\ref{virial4}) and (\ref{virial9}), we obtain for graphene
\begin{eqnarray}
D\mathcal{P}\Omega=\oint\limits_{\partial\Omega}d\mathbf{s}\cdot\mathbf{J}[G]-p_\mathrm{c}\frac{\partial E}{\partial
p_\mathrm{c}}.\label{p3}
\end{eqnarray}
To draw connection between the boundary term in (\ref{p3}) and the physical pressure caused by electron collisions with
the boundary, as in Section~\ref{local_pressure_massive}, we again need to consider the boundary relationship for the
wave function.

\subsection{Boundary relationship and local pressure}\label{boundary_Dirac}
Due to the specific form of boundary condition $M\psi=\psi$, the boundary relationship for massless Dirac electrons
will be different from (\ref{boundary1}) for massive electrons. We can consider $M\psi-\psi$ as the four-component
function satisfying the Dirichlet condition on the boundary, so Eq.~(\ref{boundary1}) with the replacement
$\psi\rightarrow M\psi-\psi$ can be applied in this case:
\begin{eqnarray}
M\delta\psi=\delta\psi-\delta\mathbf{r}_0\cdot M\nabla\psi+\delta\mathbf{r}_0\cdot\nabla\psi
+\mathcal{O}([\delta\mathbf{r}_0]^2).\label{boundary3}
\end{eqnarray}
The matrix $M$ should be unitary, $M^+=M^{-1}$, and anticommuting with the normal component
$\Sigma_n\equiv\boldsymbol\Sigma\cdot\mathbf{n}$ of the probability current operator, $\{M,\Sigma_n\}=0$, to ensure
that the particles do not cross the boundary, i.e. $\mathbf{J}[1]\cdot\mathbf{n}=v_\mathrm{F}\psi^+\Sigma_n\psi=0$
\cite{Akhmerov,McCann}. Using these properties of $M$, we get
$\psi^+\Sigma_n\delta\psi=\psi^+M\Sigma_n\delta\psi=-\psi^+\Sigma_nM\delta\psi$. Applying (\ref{boundary3}) and again
the condition $\{M,\Sigma_n\}=0$, we obtain
\begin{eqnarray}
\psi^+\Sigma_n\delta\psi=-\delta\mathbf{r}_0\cdot\psi^+\Sigma_n\nabla\psi+\mathcal{O}([\delta\mathbf{r}_0]^2).
\label{boundary4}
\end{eqnarray}
This is the counterpart of the boundary relationship (\ref{boundary1}) for massless Dirac electrons.

If the single-particle stationary state $\psi$ is unaffected by the momentum cutoff both before and after the boundary
perturbation, then the energy change of this state can be found by using (\ref{curr1}), (\ref{delta_E}), (\ref{curr4}),
and Gauss theorem:
\begin{eqnarray}
\delta E=-i\hbar v_\mathrm{F}\oint\limits_{\partial\Omega}ds\:\psi^+\Sigma_n\delta\psi.\label{delta_E_Dirac}
\end{eqnarray}
Using (\ref{boundary4}) and introducing the kinetic stress tensor for massless Dirac particles,
\begin{eqnarray}
T_{\mu\nu}\equiv J_\nu[p_\mu]=-i\hbar v_\mathrm{F}\psi^+\nabla_\mu\Sigma_\nu\psi,\label{T_Dirac}
\end{eqnarray}
we get the formulas, which are fully analogous to (\ref{ploc0})--(\ref{ploc1}). In this derivation we did not used any
specific form of $M$, requiring only the absence of particle flux through the boundary, thus its results should be
applicable to any bounded system of massless Dirac electrons.

\subsection{Thermodynamic pressure: kinetic and anomalous parts}
Generalization of the results of Section~\ref{boundary_Dirac} for a many-body system should be done with caution
because of the presence of the momentum cutoff. A generic many-body wave function $\Psi_n$ can be presented as a sum of
factorized wave functions
\begin{eqnarray}
\Psi_n(\mathbf{r}_1\ldots\mathbf{r}_N)=\sum_k C_{n}^{(k)}\psi_1^{(k)}(\mathbf{r}_1)\ldots\psi_N^{(k)}(\mathbf{r}_N),
\label{Psi_n_Dirac}
\end{eqnarray}
where each $\psi_i^{(k)}$ satisfies the boundary conditions and does not need to be an eigenfunction of the
Hamiltonian. When the boundary is perturbed, several transformation occur with this function: first, the
single-particle wave functions $\psi_i^{(k)}$, which satisfy the momentum cutoff both before and after perturbation,
are changed by the values $\delta\psi_i^{(k)}$ obeying (\ref{boundary4}) on the boundary. Second, some terms in
(\ref{Psi_n_Dirac}) disappear because one or several of $\psi_i^{(k)}$ in these terms cease to satisfy the momentum
cutoff condition, and some new terms with $\psi_i^{(k)}$ satisfying the momentum cutoff condition after the
perturbation can appear instead of the disappeared ones. Consequently, the perturbation
$\delta\Psi_n=\delta_\mathrm{def}\Psi_n+\delta_\mathrm{c}\Psi_n$ can be presented as a sum of
\begin{eqnarray}
\delta_\mathrm{def}\Psi_n(\mathbf{r}_1\ldots\mathbf{r}_N)=\sum_k
C_{n}^{(k)}\nonumber\\
\times\left[\delta\psi_1^{(k)}(\mathbf{r}_1)\ldots\psi_N^{(k)}(\mathbf{r}_N)+\ldots\right.\nonumber\\
\left.+\psi_1^{(k)}(\mathbf{r}_1)\ldots\delta\psi_N^{(k)}(\mathbf{r}_N)\right],\label{delta_def}
\end{eqnarray}
which is caused by deformations of the single-particle functions $\psi_i^{(k)}$, and $\delta_\mathrm{c}\Psi_n$, which
is caused by momentum cutoff.

Two contributions to $\delta\Psi_n$ will result, through (\ref{delta_E}), in two parts of the energy response $\delta
E$ to the volume change $\delta\Omega$, and, correspondingly, in two parts of the pressure. Assuming a uniform dilation
of the system boundary $\delta\mathbf{r}_0=(\delta R/R)\mathbf{r}_0$ and applying (\ref{boundary4}) for each
$\delta\psi_i^{(k)}$ in (\ref{delta_def}), we obtain the energy change due to single-particle wave function
deformations: $\delta_\mathrm{def}E=-(\delta R/R)\oint_{\partial\Omega}d\mathbf{s}\cdot\mathbf{J}[G]$, which is
analogous to that for massive electrons. We can call the corresponding part of the thermodynamic pressure
$\mathcal{P}=\mathcal{P}_\mathrm{kin}+\mathcal{P}_\mathrm{anom}$ as kinetic pressure
\begin{eqnarray}
\mathcal{P}_\mathrm{kin}=\displaystyle\frac1{D\Omega}\oint\limits_{\partial\Omega}d\mathbf{s}\cdot\mathbf{J}[G]
\nonumber\\ =\frac1{D\Omega}\langle H_\mathrm{kin}+V_\mathrm{int}-\mathbf{r}\cdot\nabla U_\mathrm{ext}\rangle,
\label{P_kin}
\end{eqnarray}
which is caused, from the physical point of view, by a transfer of momentum to the boundary during electron
reflections. The second part of $\mathcal{P}$, according to (\ref{p3}), is equal to
\begin{eqnarray}
\mathcal{P}_\mathrm{anom}=-\frac{p_\mathrm{c}}{D\Omega}\frac{\partial E}{\partial p_\mathrm{c}}.\label{P_anom}
\end{eqnarray}
and can be called anomalous pressure. It is caused by redistribution of electron states during a change of $\Omega$ due
to the presence of the cutoff, which provide an additional contribution to the energy change.

Thus, in contrast to massive electrons, where the thermodynamic and kinetic pressures are equal \cite{Marc}, in the
case of massless Dirac electrons they differ by $\mathcal{P}_\mathrm{anom}$, and the formulas (\ref{ploc1}),
(\ref{ploc_integral}) and (\ref{p_bulk_theorem}) are applicable only to $\mathcal{P}_\mathrm{kin}$.

We should note that the breaking of the equality between $\mathcal{P}$ and $\mathcal{P}_\mathrm{kin}$ is not restricted
to the Dirac model, but can appear in any solid with a filled valence band. The previously considered Dirac model is
the important example of such systems because the valence band cutoff is not just a formal tool used to bound the
system energy from below, but has observable consequences, e.~g., logarithmic renormalization of the Fermi velocity of
electrons in graphene in induced by Coulomb interaction \cite{CastroNeto}. Also, the Dirac model is convenient to work
with because it uniformly describes both valence and conduction bands using just a single parameter, $v_\mathrm{F}$.
However we can extend our consideration to the system with quadratic electron dispersion and energy gap between
conduction and valence bands, which is the simplest model of semiconductor or insulator. In \ref{Appendix_C} we
calculate kinetic and anomalous parts of the pressure for this model.

\section{Examples}\label{sec_examples}

\subsection{Free noninteracting electrons in thermodynamic limit}\label{sec_noninteracting}
\begin{figure}[t]
\begin{center}
\resizebox{1.\columnwidth}{!}{\includegraphics{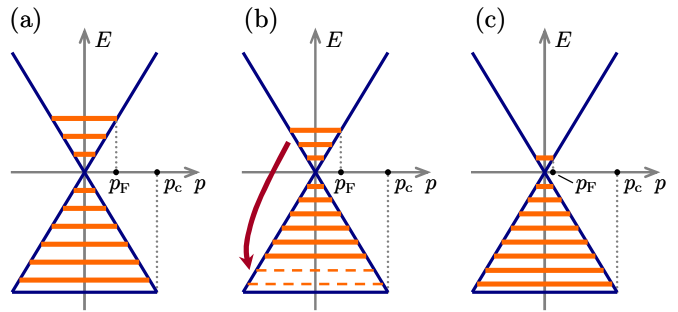}}
\end{center}
\caption{\label{Fig2}(a) Initial state of the electron-doped system of massless Dirac electrons occupying the states
from the Fermi level down to the momentum cutoff. (b) After adiabatic decrease of the electron momenta due to increase
of the system size, new unoccupied single-particle states marked by dashed lines appear in the bottom. (c) After
electron transfers from the Fermi level to the unoccupied states the ground state of the many-body system is restored.}
\end{figure}

Let us consider the simplest example of free two-dimensional noninteracting electrons occupying the single-particle
states $\psi_{\mathbf{p}\gamma}(\mathbf{r})=e^{i\mathbf{p}\cdot\mathbf{r}/\hbar}(1,\gamma
e^{i\varphi_\mathbf{p}})^T/\sqrt{2\Omega}$ with the energies $\epsilon_{\mathbf{p}\gamma}=\gamma v_\mathrm{F}p$ in the
conduction ($\gamma=+1$) and valence ($\gamma=-1$) bands in the $\mathbf{K}$ valley of graphene. In the case of
electron doping, the states in conduction and valence bands are filled up to the Fermi $p_\mathrm{F}$ and cutoff
$p_\mathrm{c}$ momenta respectively [see Fig.~\ref{Fig2}(a)]. Without attributing the exact form of boundary
conditions, we can reasonably assume that in the sample of the linear size $R$ the momenta $\mathbf{p}$ are quantized
in the units of $2\pi\hbar/R$. This neglect of the boundary behavior should be justified in the thermodynamic limit of
a large system. The total energy $E$ and number $N$ of electrons can be calculated in the thermodynamic limit by
transforming sums over momenta into integrals:
\begin{eqnarray}
E=g\sum_{\mathbf{p}\gamma}\epsilon_{\mathbf{p}\gamma}=\frac{g\Omega
v_\mathrm{F}}{6\pi\hbar^2}(p_\mathrm{F}^3-p_\mathrm{c}^3),\\
N=g\sum_{\mathbf{p}\gamma}1=\frac{g\Omega}{4\pi\hbar^2}(s_\mu p_\mathrm{F}^2+p_\mathrm{c}^2),\label{n2d}
\end{eqnarray}
where the sign $s_\mu$ of the chemical potential distinguishes the cases of electron ($s_\mu=+1$) or hole ($s_\mu=-1$)
doping, and $g=4$ is the degeneracy over valleys and spin projections. To calculate the pressure
$\mathcal{P}=-(\partial E/\partial\Omega)_N$, we need to consider simultaneous changes of $\Omega\propto R^2$ and
$p_\mathrm{F}$ which preserve $N$, and the result is:
\begin{eqnarray}
\mathcal{P}=\frac{gv_\mathrm{F}}{12\pi\hbar^2}(p_\mathrm{F}^3+3s_\mu
p_\mathrm{F}p_\mathrm{c}^2+2p_\mathrm{c}^3).\label{P2}
\end{eqnarray}
On the other hand, we can calculate kinetic and anomalous parts of $\mathcal{P}$ separately. Using (\ref{P_kin}), we
find that each electron state contributes $\epsilon_{\mathbf{p}\gamma}/2\Omega$ to $\mathcal{P}_\mathrm{kin}$, so
\begin{eqnarray}
\mathcal{P}_\mathrm{kin}=\frac{gv_\mathrm{F}}{12\pi\hbar^2}(p_\mathrm{F}^3-p_\mathrm{c}^3).\label{P_kin2}
\end{eqnarray}
Note that electrons in the valence band provide large negative contribution to $\mathcal{P}_\mathrm{kin}$ because they
have negative group velocity that implies negative momentum transfer to the boundary on collisions. The anomalous part
of pressure can by calculated from (\ref{P_anom}) with taking into account that $p_\mathrm{c}$ and $p_\mathrm{F}$
should change simultaneously to preserve $N$:
\begin{eqnarray}
\mathcal{P}_\mathrm{anom}=\frac{gv_\mathrm{F}}{4\pi\hbar^2}p_\mathrm{c}^2(s_\mu p_\mathrm{F}+
p_\mathrm{c}).\label{P_anom2}
\end{eqnarray}
The sum of (\ref{P_kin2}) and (\ref{P_anom2}) gives (\ref{P2}) in agreement with the generalized virial theorem.

The origin of $\mathcal{P}_\mathrm{anom}$ can be traced by looking at Fig.~\ref{Fig2} showing the case $s_\mu=+1$. When
$R$ is slightly increased, the momentum quantization interval decreases, that shifts the energies of the occupied
states closer to the Dirac point and results in the total energy change responsible for the kinetic pressure: $\delta
E=-\mathcal{P}_\mathrm{kin}\delta\Omega$. Since $p_\mathrm{c}\gg p_\mathrm{F}$, this $\delta E$ will be positive, hence
$\mathcal{P}_\mathrm{kin}<0$. However, due to the same decrease of the quantization interval, new unoccupied states
appear at the bottom of the valence band [Fig.~\ref{Fig2}(b)]. To maintain the ground state, the system should fill
these states with electrons taken from the Fermi level [Fig.~\ref{Fig2}(c)]. The number of transferred electrons is
proportional to $p_\mathrm{c}^2$ and their energy changes are $-v_\mathrm{F}(p_\mathrm{F}+s_\mu p_\mathrm{c})$, thus we
obtain the additional negative contribution $\delta E=-\mathcal{P}_\mathrm{kin}\delta\Omega$ to the energy change,
giving rise to the positive anomalous pressure (\ref{P_anom2}).

The nonlinear corrections to the dispersion away from the Dirac point affect the pressure quantitatively, as is shown
in more detail in \ref{Appendix_D}, but preserve the general picture of two physically different contributions to the
thermodynamic pressure.

\subsection{Interacting uniform electron gas in graphene}\label{sec_uniform}
In the case of spatially uniform electron gas in graphene with the Coulomb interaction, the expressions
(\ref{P2})--(\ref{P_anom2}) acquire interaction-induced corrections. Writing the energy
$E=\Omega\epsilon(n,p_\mathrm{c})$ and electron number $N=\Omega n$ in terms of energy $\epsilon$ and electron $n$
densities, we can rewrite the generalized virial theorem (\ref{virial9}) in the form
\begin{eqnarray}
-3\epsilon+2n\left(\frac{\partial\epsilon}{\partial n}\right)_{p_\mathrm{c}}+
p_\mathrm{c}\left(\frac{\partial\epsilon}{\partial p_\mathrm{c}}\right)_n=0,\label{virial_uniform1}
\end{eqnarray}
whereas the pressure components (\ref{P_kin})--(\ref{P_anom}) and the total pressure become
\begin{eqnarray}
\mathcal{P}_\mathrm{kin}=\frac12\epsilon,\quad
\mathcal{P}_\mathrm{anom}=-\frac12p_\mathrm{c}\left(\frac{\partial\epsilon}{\partial p_\mathrm{c}}\right)_n,\nonumber\\
\mathcal{P}=-\epsilon+n\left(\frac{\partial\epsilon}{\partial n}\right)_{p_\mathrm{c}}.\label{pressures_uniform1}
\end{eqnarray}
The theorem (\ref{virial_uniform1}) is equivalent to $\mathcal{P}=\mathcal{P}_\mathrm{kin}+\mathcal{P}_\mathrm{anom}$.

In the low-energy physics of graphene \cite{CastroNeto} only the properties of electron gas at low doping levels
$p_\mathrm{F}\ll p_\mathrm{c}$ are observable. To describe the properties of such low-doping Dirac electron gas, we
introduce the regularized energy and electron densities, obtained after subtraction of the valence band contribution:
$\epsilon_\mathrm{r}=\epsilon-\epsilon_0-\mu_0n_\mathrm{r}$, $n_\mathrm{r}=n-n_0=s_\mu gp_\mathrm{F}^2/4\pi\hbar^2$,
where $n_0=gp_\mathrm{c}^2/4\pi\hbar^2$ is the density of electrons in the filled valence band,
$\epsilon_0=\epsilon(n_0,p_\mathrm{c})$ is the energy density of the filled valence band,
$\mu_0=(\partial\epsilon/\partial n)_{p_\mathrm{c}}|_{n=n_0}$ is the chemical potential of the electron gas at
$n_\mathrm{r}=0$. Thus at $n_\mathrm{r}=0$, when the Fermi level is located in the Dirac point, we have
$\epsilon_\mathrm{r}=0$ and the regularized chemical potential $(\partial\epsilon_\mathrm{r}/\partial
n_\mathrm{r})_{p_\mathrm{c}}=0$. Switching from $\epsilon(n,p_\mathrm{c})$ to
$\epsilon_\mathrm{r}(n_\mathrm{r},p_\mathrm{c})$ and taking into account that, according to dimensionality,
$\varepsilon_0\propto p_\mathrm{c}^3$ and $\mu_0\propto p_\mathrm{c}$, we obtain the regularized version of the
generalized virial theorem, previously discussed in Ref.~\cite{Sokolik}:
\begin{eqnarray}
-3\epsilon_\mathrm{r}+2n_\mathrm{r}\left(\frac{\partial\epsilon_\mathrm{r}}{\partial n_\mathrm{r}}\right)_{p_\mathrm{c}}+
p_\mathrm{c}\left(\frac{\partial\epsilon_\mathrm{r}}{\partial p_\mathrm{c}}\right)_{n_\mathrm{r}}=0.\label{virial_uniform2}
\end{eqnarray}
Similarly to (\ref{pressures_uniform1}), we introduce the regularized kinetic, anomalous and total pressures,
\begin{eqnarray}
\mathcal{P}_\mathrm{r,kin}=\frac12\epsilon_\mathrm{r},\quad
\mathcal{P}_\mathrm{r,anom}=-\frac12p_\mathrm{c}\left(\frac{\partial\epsilon_\mathrm{r}}{\partial
p_\mathrm{c}}\right)_{n_\mathrm{r}},\nonumber\\
\mathcal{P}_\mathrm{r}=-\epsilon_\mathrm{r}+n_\mathrm{r}\left(\frac{\partial\epsilon_\mathrm{r}}{\partial
n_\mathrm{r}}\right)_{p_\mathrm{c}},\label{pressures_uniform2}
\end{eqnarray}
so the theorem (\ref{virial_uniform2}) states
$\mathcal{P}_\mathrm{r}=\mathcal{P}_\mathrm{r,kin}+\mathcal{P}_\mathrm{r,anom}$. Despite the similarity of
(\ref{virial_uniform1})--(\ref{pressures_uniform1}) and (\ref{virial_uniform2})--(\ref{pressures_uniform2}), the
quantities entering these formulas are very different in value and dependencies on system parameters.

To reveal the meaning of $\mathcal{P}_\mathrm{r,kin}$ and $\mathcal{P}_\mathrm{r,anom}$, we can use the scaling form of
energy density: $\epsilon_\mathrm{r}=\epsilon_\mathrm{r}^{(0)}f(\Lambda,r_\mathrm{s})$, where
$\epsilon_\mathrm{r}^{(0)}=gv_\mathrm{F}p_\mathrm{F}^3/6\pi\hbar^2$ is the regularized energy density of noninteracting
gas, $\Lambda=p_\mathrm{c}/p_\mathrm{F}$ is the dimensionless cutoff momentum, and $r_\mathrm{s}=e^2/\varepsilon\hbar
v_\mathrm{F}$ is the Coulomb interaction scale (``fine structure constant'') for graphene, $\varepsilon$ is the
dielectric constant of surrounding medium. In terms of $f$, we obtain
$\mathcal{P}_\mathrm{r,anom}=-(\epsilon_\mathrm{r}^{(0)}/2)\Lambda(\partial f/\partial\Lambda)$. As we see,
$\mathcal{P}_\mathrm{r,anom}$ is caused by the cutoff dependence of the energy, which appears only in the presence of
Coulomb interaction, because $f=1$ and $\mathcal{P}_\mathrm{r,anom}=0$ at $r_\mathrm{s}=0$ (in contrast to
$\mathcal{P}_\mathrm{anom}$, which is nonzero even at $r_\mathrm{s}=0$, see (\ref{P_anom2})). As shown in
Ref.~\cite{Sokolik}, $\mathcal{P}_\mathrm{r,anom}$ measures the extent of scale invariance breaking due to the cutoff
in the theorem (\ref{virial_uniform2}). So, even after subtraction of the valence band contributions to energy and
chemical potential, the electron gas properties continue to depend on the cutoff momentum $p_\mathrm{c}$ in the
presence of interaction.

As a specific example, we can take the Hartree-Fock approximation, in which function $f$ can be expanded at large
$\Lambda$ with sufficient accuracy as \cite{Lozovik,Peres}
\begin{eqnarray}
f=1+r_\mathrm{s}\left\{\frac14\ln\Lambda+\frac12\ln2-\frac1{24}-\frac{2\mathcal{C}+1}{2\pi}\right.\nonumber\\
\left.+\frac{3s_\mu}{32\Lambda}+\mathcal{O}\left(\frac1{\Lambda^2}\right)\right\},\label{f_HF}
\end{eqnarray}
where $\mathcal{C}\approx0.916$ is Catalan's constant. In more accurate random-phase approximation
\cite{Lozovik,Barlas}, the coefficients at $\ln\Lambda$ and $\Lambda^{-n}$ ($n\geqslant0$) acquire additional nonlinear
$r_\mathrm{s}$ dependencies. If, generally, $f=1+A(r_\mathrm{s})\ln\Lambda+\mathcal{O}(1/\Lambda)$ ($A=r_\mathrm{s}/4$
in the Hartree-Fock approximation), then the leading $\ln\Lambda$ term is related to the quantity
\begin{eqnarray}
K=\Lambda\frac{\partial f}{\partial\Lambda}=\frac{p_\mathrm{c}}{\epsilon_\mathrm{r}^{(0)}}\left(\frac{\partial\epsilon_\mathrm{r}}{\partial p_\mathrm{c}}\right)_{n_\mathrm{r}}=-\frac{2\mathcal{P}_\mathrm{r,anom}}{\epsilon_\mathrm{r}^{(0)}},
\end{eqnarray}
which was discussed in Ref.~\cite{Sokolik} and can be evaluated from experimental data on graphene electron
compressibility or quantum capacitance. At $\Lambda\rightarrow\infty$, we have
$K=A(r_\mathrm{s})+\mathcal{O}(1/\Lambda)$. The numerical calculations show that $K$ is nearly constant in the range of
doping levels of graphene accessible by using the electric field effect \cite{Sokolik}. Thus the regularized anomalous
pressure $\mathcal{P}_\mathrm{r,anom}$ can be related to experimental data and to the logarithmic term in $f$. Note
that, as seen from (\ref{f_HF}), $\mathcal{P}_\mathrm{r,kin}$ is also changed in the presence of interaction due to
renormalization of electron Fermi velocity \cite{CastroNeto}.

\begin{figure*}[t]
\begin{center}
\resizebox{0.8\textwidth}{!}{\includegraphics{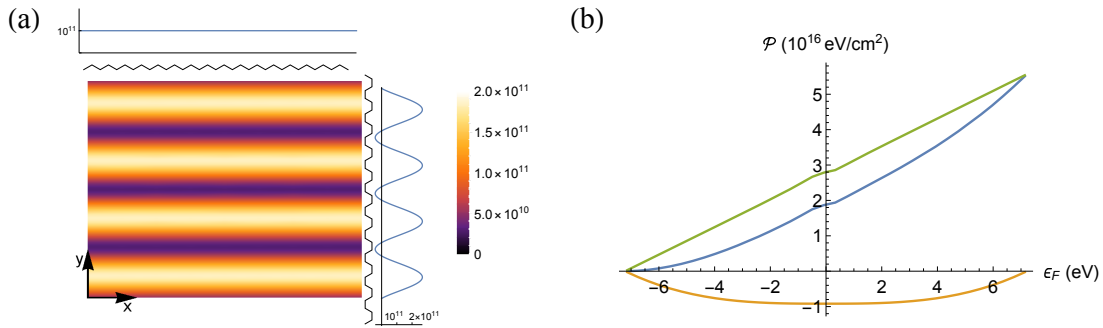}}
\end{center}
\caption{\label{Fig3}(a) Bulk and boundary kinetic pressure (in the units of eV/cm) of the single-particle state with
quantum numbers $m=5$, $n=4$, $\epsilon>0$ of massless Dirac electron in a rectangular graphene flake of dimensions
$L_x=20$ nm, $L_y=16$ nm. (b) Kinetic (orange line, bottom), anomalous (green line, top), and total thermodynamic (blue
line, middle) pressure of Dirac electrons in a rectangular graphene flake as a function of the Fermi energy. Dimensions
$L_{x,y}$ are the same as in (a).}
\end{figure*}

\begin{figure}[b]
\begin{center}
\resizebox{0.8\columnwidth}{!}{\includegraphics{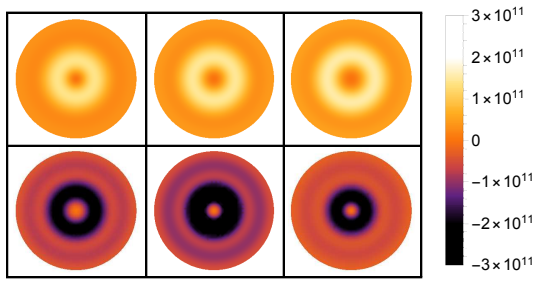}}
\end{center}
\caption{\label{Fig4} Distribution of local bulk pressure (in $\mbox{eV/cm}^2$, without spin and valley degeneracies)
for single-particle states in the circular graphene flake with radius $R=10\,\mbox{nm}$ at quantum numbers $j=3/2$,
$n=2$, $\gamma=\pm1$ in, respectively, the top and the bottom row, and for the Coulomb impurity dimensionless charges
$\tilde{g}=-0.4,0,0.4$ in, respectively, the left, middle, and right columns.}
\end{figure}

Both regularized and nonregularized thermodynamic pressures are connected with the observable quantum capacitance of
graphene $C_\mathrm{Q}$ \cite{Lozovik}:
\begin{eqnarray}
e^2C_\mathrm{Q}^{-1}=\frac1n\left(\frac{\partial\mathcal{P}}{\partial n}\right)_{p_\mathrm{c}}=\frac1{n_\mathrm{r}}\left(\frac{\partial\mathcal{P_\mathrm{r}}}{\partial n_\mathrm{r}}\right)_{p_\mathrm{c}}.
\end{eqnarray}
In principle, we can calculate the total, unregularized pressures $\mathcal{P}$, $\mathcal{P}_\mathrm{kin}$, and
$\mathcal{P}_\mathrm{anom}$ using the known dependence $C_\mathrm{Q}^{-1}(n_\mathrm{r})$:
\begin{eqnarray}
\mathcal{P}=-\epsilon_0+\mu_0n_0+e^2\int\limits_0^{n-n_0}(n_0+n_r')C_\mathrm{Q}^{-1}(n_\mathrm{r}')\:dn_\mathrm{r}',\label{P_uniform}\\
\mathcal{P}_\mathrm{kin}=\frac12\epsilon_0+\frac12\mu_0(n-n_0)\nonumber\\+\frac{e^2}2\int\limits_0^{n-n_0}(n-n_0-n_r')C_\mathrm{Q}^{-1}(n_\mathrm{r}')\:dn_\mathrm{r}',\label{P_dyn_uniform}
\end{eqnarray}
and $\mathcal{P}_\mathrm{anom}=\mathcal{P}-\mathcal{P}_\mathrm{kin}$. Using the power series expansion of
$C_\mathrm{Q}^{-1}$ for the interacting system, similar to (\ref{f_HF}), we can perform the integrations analytically
and obtain the interaction-induced corrections to the unregularized pressures, caused by corresponding corrections to
$C_\mathrm{Q}^{-1}$ studied in Ref.~\cite{Lozovik}. However the formulas (\ref{P_uniform})--(\ref{P_dyn_uniform})
contain the parameters $n_0$, $\epsilon_0$, and $\mu_0$ of the filled valence band, which can be the sources of
additional interaction-induced corrections. In estimating these parameters, we also need to take into account
deviations from the linear dispersion at large electron momenta (\ref{Appendix_D}).

\subsection{Rectangular graphene flake}\label{sec:rect}
Here we consider the single-particle states of massless Dirac electrons in a rectangular graphene sample with zigzag
horizontal edges and armchair vertical edges, imposing the corresponding boundary conditions:
$\psi_{A\mathbf{K}}=\psi_{A\mathbf{K}'}=0$ at the bottom edge, $\psi_{B\mathbf{K}}=\psi_{B\mathbf{K}'}=0$ at the top
edge, $\psi_{A\mathbf{K}}+\psi_{A\mathbf{K}'}=\psi_{B\mathbf{K}}+\psi_{B\mathbf{K}'}=0$ on the left edge, $e^{2\pi\nu
i}\psi_{A\mathbf{K}}+\psi_{A\mathbf{K}'}=e^{2\pi\nu i}\psi_{B\mathbf{K}}+\psi_{B\mathbf{K}'}=0$ on the right edge,
where $\nu=\pm2/3$ or 0 depending on the atomic-scale details \cite{Brey,Akhmerov}. Hereafter we set $\hbar\equiv1$,
$v_\mathrm{F}\equiv1$ in the formulas; in numerical calculations, we take $v_\mathrm{F}=10^6\,\mbox{m/s}$ and such
cutoff $p_\mathrm{c}$ that the filled valence band has two electrons per unit cell of graphene, which corresponds to
the cutoff energy around 7.2 eV. In a $L_x\times L_y$ rectangle, the (not normalized) eigenstates $\psi= (
e^{ik_xx}\sin k_ny $, $\mp(-1)^ne^{ik_xx}\sin k_n(L_y-y)$, $\pm(-1)^ne^{-ik_xx}\sin k_n(L_y-y) $, $-e^{-ik_xx}\sin k_ny
)^T$ with energies $\epsilon=\pm\sqrt{k_x^2+k_n^2}$ are determined by quantum numbers $m\in\mathbb{Z}$ and $n$, where $
k_x = (-\frac23 \nu + m)\pi/L_x$ (we take $\nu=0$), and $k_n$ is the $n$th positive root of the equation $k_n=-k_x \tan
k_nL_y$. The local pressure on the boundaries as defined in (\ref{ploc1}) is constant at the zigzag edges and
oscillates along the armchair ones. These oscillations occur because the zigzag edges preserve the valley of the
incident wave so in the direction normal to them a standing wave pattern is formed, while the armchair edges change the
valley so there is no interference between incident and reflected waves. Besides the local pressure on the boundaries,
we calculate the local bulk pressure (\ref{p_bulk}), which equals simply $\frac12\epsilon\psi^+\psi$. It is constant
over the $x$ direction and oscillates over the $y$ direction. The example of a typical pressure distribution for one
single-particle state is shown in Fig.~\ref{Fig3}(a). If we consider the pressure distribution of a many-body system
with many different states occupied, the oscillations of the pressure disappear, but the feature of zero boundary
pressure in the $x$ direction at the angles of the flake is preserved.

\begin{figure*}[t]
\begin{center}
\resizebox{0.8\textwidth}{!}{\includegraphics{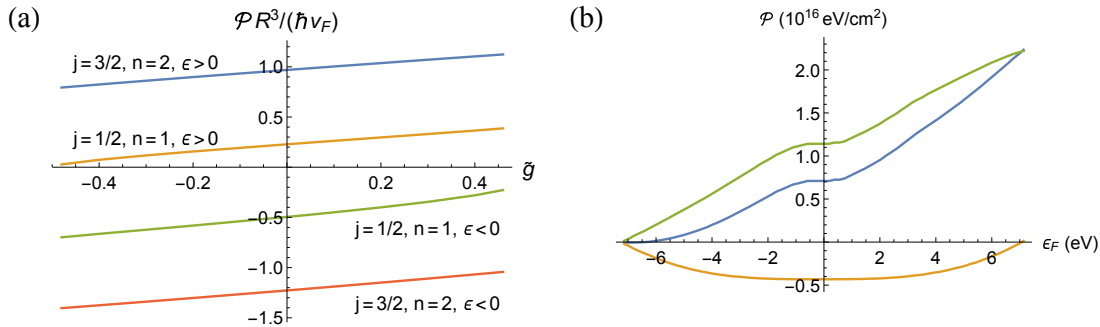}}
\end{center}
\caption{\label{Fig5}(a) The dimensionless kinetic pressure on the boundary, given by the normal momentum current, in
the circular flake of radius $R$ with Coulomb impurity as a function of the Coulomb potential parameter for various
quantum numbers. (b) Kinetic (orange line, bottom), anomalous (green line, top), and total thermodynamic (blue line,
middle) pressure of Dirac electrons in a circular graphene flake ($R=4$ nm, $\tilde{g}=-0.4$) as a function of the
Fermi energy.}
\end{figure*}

To calculate the kinetic pressure of the many-body system according to (\ref{P_kin}), which in the noninteracting case
reduces to $\mathcal{P}_\mathrm{kin}=E/2L_xL_y$, and the anomalous pressure (\ref{P_anom}), we need to consider the
total energy $E$ as a function of $p_\mathrm{c}$ with a constant $N$. Due to discrete nature of the energy spectrum,
the derivative $(\partial E/p_\mathrm{c})_N$ is the sum of Dirac delta functions, so they should be smoothened
(replaced by Lorentzians in our calculations) to get sensible result for $\mathcal{P}_\mathrm{anom}$. Both
contributions to the pressure are shown in Fig.~\ref{Fig3}(b). In the region of large momenta, where the energy levels
are spaced densely enough, the picture is expectedly very similar to the case of free Dirac electron gas, described by
Eqs. (\ref{P_kin2})--(\ref{P_anom2}). Note that the total thermodynamic pressure is positive everywhere.

\subsection{Circular graphene flake in Coulomb potential}
For a circular flake, we impose the infinite mass boundary condition $\psi_{B}=ie^{i\phi}\psi_A$, where $\phi$ is the
polar angle of the direction normal to the boundary \cite{Berry}, which decouple the valleys and allows considering
only a single valley. To study the effects of the external potential in the circular geometry, we assume the presence
of a Coulomb impurity in the center. Without external potential, the solutions of the Dirac equation are given by
Bessel functions; for a subcritical Coulomb potential $-Ze^2/r\equiv \tilde{g}\hbar v_\mathrm{F}/r$, $|\tilde{g}|<1/2$,
the solutions are given by the Coulomb wave functions \cite{Pereira}. The quantum numbers defining a solution are the
half-integer angular momentum $j$, the radial quantum number $n\in\mathbb{N}$, and the energy sign $\gamma$. The
electron-hole symmetry of the system requires $\epsilon_{\gamma,j,n}(\tilde{g})=-\epsilon_{-\gamma,-j,n}(-\tilde{g})$.
The local pressure on the boundary (\ref{ploc1}) is uniform due to the circular symmetry of the system. The
distributions of the local bulk pressure (\ref{p_bulk}), which now includes the contribution of the external Coulomb
force, are shown in Fig.~\ref{Fig4} for several single-particle states. As shown in Fig.~\ref{Fig5}(a), the states with
the same quantum numbers have higher quantum pressure (both the total kinetic pressure and the local boundary pressure)
at higher values of the Coulomb potential parameter $\tilde{g}$ in the agreement with Eq.~(\ref{P_kin}): the repulsive
potential increases the pressure by pushing the electrons towards the boundary, while the attractive potential
decreases the pressure by pulling the electrons to the center.

Fig.~\ref{Fig5}(b) shows the kinetic, anomalous (smoothened with Lorentzians), and total pressure of Dirac electrons in
a circular flake. In comparison with Fig.~\ref{Fig3}(b), here the smaller size of the flake leads to visible deviations
of the anomalous pressure from the linear trend (\ref{P_anom2}) near the Dirac point, but the agreement with
thermodynamic limit (\ref{P_kin2})--(\ref{P_anom2}) is restored at large Fermi momenta.

\section{Discussion}\label{sec_conclusions}
Using the generalized continuity equation and scaling transformations, we derived and analyzed the virial and
Hellmann--Feynman theorems for single- and many-electron systems with taking into account the presence of system
boundaries. The boundary conditions imposed on the wave function make the Hamiltonian generally non-Hermitian, which
results in appearance of additional terms in (\ref{virial2}), (\ref{HF2}) in the form of boundary integrals. We start
with the case of massive electrons and analyze the thermodynamic pressure as a response of a system energy on small
volume changes and relate the pressure to the boundary term in the virial theorem (\ref{virial7})--(\ref{virial8})
using the boundary relationships (\ref{boundary1})--(\ref{boundary2}). Besides, we find the local pressure as a
response (\ref{ploc2}) of the energy on local deformations of the boundary and connect it with the kinetic part of the
stress tensor (\ref{stress_massive}), (\ref{ploc1}). The formulas are first derived for a single-particle system and
then generalized for a many-body system in Section~\ref{sec_manybody}. While the most of these relationships for
massive electrons can be found elsewhere \cite{Marc,Ziesche,Argyres,Abad,Esteve,Cottrell,Fernandez,Bobrov,Srebrenik},
we presented them for the sake of completeness. The connection (\ref{ploc2})--(\ref{ploc1}) between energy change and
boundary perturbations is also known in the boundary perturbation theory of the boundary-value problems \cite{Henry}.

For massless Dirac electrons in a solid the similar formulas are different in some aspects because of the linear
dispersion, different forms of boundary conditions for a wave function and due to the presence of momentum cutoff in
the valence band. The latter results in appearance of the cutoff-induced term in the generalized virial theorem
(\ref{virial9}), and the thermodynamic pressure (\ref{p3}) turned out to consist of two parts. The first part is the
kinetic pressure (\ref{P_kin}), which is just a sum of responses of single-particle energies weighted with their
occupation numbers. Using the boundary relationships (\ref{boundary3}), (\ref{boundary4}) for massless Dirac electrons,
we can relate it, analogously to (\ref{ploc1}), to the kinetic stress tensor. Thus the kinetic pressure is caused by
momentum transferred by electrons to the surroundings during their reflections from the boundary. The second part is
the anomalous pressure (\ref{P_anom}), which is related to the momentum cutoff and caused by redistribution of electron
states during a volume change, as shown in Fig.~\ref{Fig2}. Note that the problem of consistency between kinetic and
thermodynamic definitions of the pressure, resolved for ordinary massive electrons \cite{Cottrell,Fernandez,Argyres}
with the help of the boundary relationships, rises again in the case of massless Dirac electrons because of the
anomalous contribution.

The example of free electrons considered in Section~\ref{sec_noninteracting} demonstrates that while the total kinetic
pressure of an electron gas in graphene is negative due to overwhelming contribution of the valence band, it is
overcompensated by the anomalous pressure, making the thermodynamic pressure positive. The examples of rectangular and
circular graphene flakes with the zigzag, armchair and infinite effective mass boundary conditions for the wave
functions demonstrate fulfilment of the general theorems.

The pressure $\mathcal{P}$ studied in this paper is related to the grand thermodynamic potential
$\Omega=-\mathcal{P}\Omega$ of the electron gas, so taking its derivatives with respect to the parameters can provide
all thermodynamic properties. The derivatives of $\mathcal{P}$ with respect to the electron density are related to such
observable quantities as electron compressibility and quantum capacitance, which were studied in graphene in the
context of interaction and disorder-induced effects \cite{Lozovik}. In Section~\ref{sec_uniform} we consider connection
of $\mathcal{P}$, $\mathcal{P}_\mathrm{kin}$, $\mathcal{P}_\mathrm{anom}$ with quantum capacitance. Our analysis of a
bounded system allows to extend these studies by including the effects of boundary conditions in small graphene flakes.
The general theorems derived here can be applied not only to graphene, but also to three-dimensional Dirac and Weyl
semimetals, which host massless Dirac electrons as well \cite{Armitage,Wehling}.

The changes of the volume of the system considered in this paper concern only electron subsystem and not the crystal lattice
itself. We analyze what happens with the electrons described by the effective Dirac equation when we move the boundary
conditions without deforming the lattice, so $p_\mathrm{c}=\mathrm{const}$ in these processes. However the other kind
of the system volume changes, when graphene is compressed or stretched as a whole, can be considered. In this case the
energy change can be related to a total mechanical stress and total mechanical compressibility of graphene. Analysis of
such graphene properties should include, besides the considered response of Dirac electrons, the responses of the core
electrons and atomic nuclei, which is beyond the scope of our paper.

The problem of breaking the equality between thermodynamic and kinetic pressures due to the anomalous contribution is not
unique for Dirac particles and can arise in any solid state system with the filled valence band. Change of electron
number in this band proportional to the change of enclosing volume requires electron transitions between valence and
conduction bands, which provide anomalous contribution to the total pressure. As the recent study \cite{Nakata}
suggests, such transitions can proceed through the Tamm states inside the energy gap.

The momentum cutoff deep in valence band of graphene, which results in appearance of the ``anomalous'' terms in the
generalized virial theorem and pressure, may be considered as an artificial construct, however in solids with massless
Dirac electrons it has real physical grounds, because valence band is indeed bounded in energy and momentum spaces. In
graphene it leads to a finite-valued logarithmic renormalization of the Fermi velocity due to Coulomb interaction
\cite{CastroNeto}. Nevertheless, more accurate analysis with going beyond the Dirac model and with taking into account
Tamm states on the boundaries can provide more insight into the problem of electron gas pressure in solids with unusual
band structure. The study of relationship between kinetic and thermodynamic pressures with taking into account other
possible anomalies can be extended to a broader context of statistical physics of confined many-particle systems.

\section*{Acknowledgments}
The work was supported by the grants No. 17-02-01134 and 18-52-00002 of the Russian Foundation of Basic Research.
Yu.E.L. was partly supported by the Program for Basic Research of the National Research University Higher School of Economics.
A.A.S. and A.D.Z. acknowledge the support from the Foundation for the Advancement of Theoretical Physics and
Mathematics ``BASIS''.

\appendix

\begin{widetext}
\section{Scaling relations for uniform system of massive electrons}\label{Appendix_A}
Consider a system of $N$ massive electrons with Coulomb interaction $V_\mathrm{int}(r)=e^2/\varepsilon r$ in the
external power-law potential $U_\mathrm{ext}(r)=U_0r^\gamma$, confined in the volume $\Omega$ by infinitely high
potential walls. This system is described by the many-body Schrodinger equation
\begin{eqnarray}
\left\{\sum_i\left(-\frac{\hbar^2\nabla_i^2}{2m}+U_0r_i^\gamma\right)+\frac12\sum_{i\neq
j}\frac{e^2}{\varepsilon|\mathbf{r}_i-\mathbf{r}_j|}\right\}\Psi=E\Psi.
\end{eqnarray}
On the boundary we impose the Dirichlet condition: $\Psi(\mathbf{r}_1\ldots\mathbf{r}_N)=0$ when $\vee i:\,
\mathbf{r}_i\in\partial\Omega$.

With the characteristic size of $\Omega$ being $R$, we can switch to the dimensionless coordinates
$\boldsymbol\rho_i=\mathbf{r}_i/R$, and the wave function is scaled as
$\Psi(\mathbf{r}_1\ldots\mathbf{r}_N)=R^{-ND/2}\tilde\Psi(\boldsymbol\rho_1\ldots\boldsymbol\rho_N)$. Introducing the
dimensionless energy $\tilde{E}=mR^2E/\hbar^2$, interaction strength $r_\mathrm{s}=e^2mR/\varepsilon\hbar^2$, and
external potential strength $\varkappa=U_0mR^{\gamma+2}/\hbar^2$, we obtain the scaled equation
\begin{eqnarray}
\left\{\sum_i\left(-\frac12\frac{\partial^2}{\partial\boldsymbol\rho_i^2}+\varkappa\rho_i^\gamma\right)+\frac12\sum_{i\neq
j}\frac{r_\mathrm{s}}{|\boldsymbol\rho_i-\boldsymbol\rho_j|}\right\}\tilde\Psi=\tilde{E}\tilde\Psi,
\end{eqnarray}
which does not depend on $R$ explicitly. The boundary conditions for $\tilde\Psi$ depend only on the shape of $\Omega$
and not on its size. As a result, we obtain the scaling forms
\begin{eqnarray}
\Psi(\mathbf{r}_1\ldots\mathbf{r}_N)=\frac1{R^{\frac{ND}2}}\tilde\Psi\left(\frac{\mathbf{r}_1}R\ldots\frac{\mathbf{r}_1}R;
\frac{e^2mR}{\varepsilon\hbar^2},\frac{U_0mR^{\gamma+2}}{\hbar^2}\right),\label{scaling1}\\
E=\frac{\hbar^2}{mR^2}\tilde{E}\left(\frac{e^2mR}{\varepsilon\hbar^2},\frac{U_0mR^{\gamma+2}}{\hbar^2}\right).
\label{scaling2}
\end{eqnarray}

Calculating derivatives of (\ref{scaling1}) and (\ref{scaling2}) with respect to $R$, we obtain the following scaling
properties of $\Psi$ and $E$:
\begin{eqnarray}
\left\{-\left(\sum_i\mathbf{r}_i\cdot\nabla_i+\frac{ND}2\right)-R\frac\partial{\partial R}
+e^2\frac\partial{\partial(e^2)}
+(\gamma+2)U_0\frac\partial{\partial U_0}\right\}\Psi=0,\label{scaling3}\\
-R\frac{\partial E}{\partial R}-2E+e^2\frac{\partial E}{\partial (e^2)}+(\gamma+2)U_0\frac{\partial E}{\partial U_0}=0.
\label{scaling4}
\end{eqnarray}
Since $e^2(\partial E/\partial (e^2))=\langle V_\mathrm{int}\rangle$ and $\gamma U_0(\partial E/\partial U_0)={\langle
\mathbf{r}\cdot\nabla U_\mathrm{ext}\rangle}$, we can immediately identify (\ref{scaling4}) as the virial theorem
(\ref{virial10}) for the case of power-law $U_\mathrm{ext}(\mathbf{r})$. Eq.~(\ref{scaling3}) should be valid in all
points of space, so if one of the $\mathbf{r}_i$ is located on the boundary, $\partial\Psi/\partial(e^2)$ and
$\partial\Psi/\partial U_0$ vanish due to the Dirichlet boundary condition, so we obtain
$(\sum_i\mathbf{r}_i\cdot\nabla_i)\Psi=-R(\partial\Psi/\partial R)$ and hence the many-body counterpart of the boundary
relationship (\ref{boundary2}).

\section{Scaling relations for uniform system of massless Dirac electrons}\label{Appendix_B}
A system of $N$ massless Dirac electrons with Coulomb interaction $V_\mathrm{int}(r)=e^2/\varepsilon r$ in the external
power-law potential $U_\mathrm{ext}(r)=U_0r^\gamma$, confined in the volume $\Omega$, is described by the many-body
Dirac equation:
\begin{eqnarray}
\left\{\sum_i\left(-i\hbar v_\mathrm{F}\boldsymbol\Sigma_i\cdot\nabla_i+U_0r_i^\gamma\right)+\frac12\sum_{i\neq
j}\frac{e^2}{\varepsilon|\mathbf{r}_i-\mathbf{r}_j|}\right\}\Psi=E\Psi.
\end{eqnarray}
Some boundary conditions of the kind $M_i\Psi=\Psi$, not specified explicitly here, are imposed on
$\Psi(\mathbf{r}_1\ldots\mathbf{r}_N)$ when $\vee i:\, \mathbf{r}_i\in\partial\Omega$. We should also impose the
momentum cutoff condition $P_{p_\mathrm{c}}\Psi=\Psi$, where the operator $P_{p_\mathrm{c}}$ of projection on the
subspace $|\mathbf{p}_i|\leq p_\mathrm{c}$ of momentum space was described in Ref.~\cite{Sokolik}.

As in \ref{Appendix_A}, we use the dimensionless coordinates $\boldsymbol\rho_i=\mathbf{r}_i/R$ and the scaled
wave function $\Psi(\mathbf{r}_1\ldots\mathbf{r}_N)=R^{-ND/2}\tilde\Psi(\boldsymbol\rho_1\ldots\boldsymbol\rho_N)$.
Introducing the dimensionless energy $\tilde{E}=RE/\hbar v_\mathrm{F}$, interaction constant
$r_\mathrm{s}=e^2/\varepsilon\hbar v_\mathrm{F}$, and the external potential strength $\varkappa=U_0R^{\gamma+1}/\hbar
v_\mathrm{F}$, we obtain the scaled Dirac equation:
\begin{eqnarray}
\left\{\sum_i\left(-i\boldsymbol\Sigma_i\cdot\frac\partial{\partial\boldsymbol\rho_i}+\varkappa\rho_i^\gamma\right)
+\frac12\sum_{i\neq j}\frac{r_\mathrm{s}}{|\boldsymbol\rho_i-\boldsymbol\rho_j|}\right\}\tilde\Psi=\tilde{E}\tilde\Psi.
\end{eqnarray}
The boundary conditions for $\tilde\Psi$ are now independent on $R$, and the cutoff condition depends only on the
dimensionless parameter $\Lambda=Rp_\mathrm{c}$. The resulting scaling forms of $\Psi$ and $E$ are
\begin{eqnarray}
\Psi(\mathbf{r}_1\ldots\mathbf{r}_N)=\frac1{R^{\frac{ND}2}}\tilde\Psi\left(\frac{\mathbf{r}_1}R\ldots\frac{\mathbf{r}_1}R;
\frac{e^2}{\varepsilon\hbar v_\mathrm{F}},\frac{U_0R^{\gamma+1}}{\hbar v_\mathrm{F}},Rp_\mathrm{c}\right),
\label{scaling5}\\
E=\frac{\hbar v_\mathrm{F}}{R}\tilde{E}\left(\frac{e^2}{\varepsilon\hbar v_\mathrm{F}},\frac{U_0R^{\gamma+1}}{\hbar
v_\mathrm{F}},Rp_\mathrm{c}\right).\label{scaling6}
\end{eqnarray}

Calculating derivatives of (\ref{scaling5}) and (\ref{scaling6}) with respect to $R$, we obtain the scaling properties:
\begin{eqnarray}
\left\{-\left(\sum_i\mathbf{r}_i\cdot\nabla_i+\frac{ND}2\right)-R\frac\partial{\partial R}
+(\gamma+1)U_0\frac\partial{\partial U_0}\right\}\Psi=0,\label{scaling7}\\
-R\frac{\partial E}{\partial R}+p_\mathrm{c}\frac{\partial E}{\partial p_\mathrm{c}}-2E+(\gamma+1)U_0\frac{\partial
E}{\partial U_0}=0. \label{scaling8}
\end{eqnarray}
Taking into account that $\gamma U_0(\partial E/\partial U_0)={\langle \mathbf{r}\cdot\nabla U_\mathrm{ext}\rangle}$ we
obtain from (\ref{scaling8}) the generalized virial theorem (\ref{virial9}). The equation (\ref{scaling7}) can be
interpreted as a counterpart of (\ref{boundary2}) for the many-body wave function subject to momentum cutoff. The
scaling analysis of a system of massless Dirac electrons can be also found in \cite{Lin}.

\section{Massive electrons in a two-band model}\label{Appendix_C}
Consider a two-band model with nonzero effective mass:
\begin{eqnarray}
H_\mathrm{kin}=\left(\Delta+\frac{p^2}{2m}\right)\sigma_z.\label{ham-mass-2band}
\end{eqnarray}
This is the simplest model description of conduction and valence bands of a semiconductor or insulator separated by the
gap $2\Delta$.

If the momentum cutoff at $p=p_\mathrm{c}$ is imposed in the valence band, it can be shown in the same way as in the
Dirac case that the anomalous pressure arises in such system, too. In particular, in the two-dimensional noninteracting
many-body system with degeneracy factor $g$, the number of particles $N$ is given by Eq.~(\ref{n2d}), while the energy,
the thermodynamic pressure, and the kinetic and the anomalous contributions to it are given by the following
expressions (the derivatives are taken at constant $N$):
\begin{eqnarray}
E&=&\frac{g\Omega}{2\pi\hbar^2}\left\{\frac{p_\mathrm{F}^4-p_\mathrm{c}^4}{8m}+
\Delta\frac{p_\mathrm{F}^2-p_\mathrm{c}^2}2\right\},\\
\mathcal{P}&=&-\frac{\partial E}{\partial\Omega}=\frac{g}{2\pi\hbar^2} \left\{\frac{(s_\mu
p_\mathrm{F}^2+p_\mathrm{c}^2)^2}{8m}+
(s_\mu+1)\Delta\frac{p_\mathrm{c}^2}2\right\},\\
\mathcal{P}_\mathrm{kin}&=&\left.\frac{E}{\Omega}\right|_{\Delta\rightarrow0}=
\frac{g}{2\pi\hbar^2}\frac{p_\mathrm{F}^4-p_\mathrm{c}^4}{8m},\\
\mathcal{P}_\mathrm{anom}&=&-\frac{p_\mathrm{c}}{2\Omega}\frac{\partial E}{\partial p_\mathrm{c}}=
\frac{g}{2\pi\hbar^2}\left\{\frac{(s_\mu p_\mathrm{F}^2+p_\mathrm{c}^2)p_\mathrm{c}^2}{4m}+
(s_\mu+1)\Delta\frac{p_\mathrm{c}^2}2\right\}.
\end{eqnarray}
Here $s_\mu=\pm1$ for, respectively, electron- and hole-doped material. Similarly to the case of noninteracting Dirac
electrons (Section \ref{sec_noninteracting}), at $p_\mathrm{c}\gg p_\mathrm{F}$ we have $\mathcal{P}_\mathrm{kin}<0$,
but the total pressure is positive due to anomalous contribution.

\section{Corrections to pressure due to dispersion nonlinearities}\label{Appendix_D}
For noninteracting Dirac electrons in graphene with perfectly linear dispersion, the kinetic and the anomalous pressure
is given by the equations (\ref{P_kin2}) and (\ref{P_anom2}). However far away from the Dirac point the dispersion has
nonlinear corrections \cite{CastroNeto}. Here we consider the corrections to the pressure from these nonlinearities.

If the nearest-neighbor tight-binding Hamiltonian for the $2p_z$ orbitals of carbon atoms in graphene is expanded near
the $\mathbf{K}$ point \cite{CastroNeto}, the first term is linear in the momentum $p$, the next term is the trigonal
warping proportional to $p^2\sin3\varphi_{\mathbf{p}}$ which provides no contribution to the pressure after integration
over the polar angle $\varphi_{\mathbf{p}}$, and the cubic term is:
$\delta\epsilon_{\mathbf{p}\gamma}^{(3)}=-(7/64)\gamma v_\mathrm{F} p^3(d/\hbar)^2$ where $d$ is the interatomic
distance. The corrections to the kinetic and anomalous pressure caused by $\delta\epsilon_{\mathbf{p}\gamma}^{(3)}$
are:
\begin{eqnarray}
\delta\mathcal{P}_\mathrm{kin}^{(3)}=\frac{21g\kappa v_\mathrm{F}}{1280\pi\hbar^2}
\left(1-\frac{p_\mathrm{F}^5}{p_\mathrm{c}^5}\right)p_\mathrm{c}^3,\quad \delta\mathcal{P}_\mathrm{anom}^{(3)}=
-\frac{7g\kappa v_\mathrm{F}}{256\pi\hbar^2}(s_\mu p_\mathrm{F}^3  + p_\mathrm{c}^3),
\end{eqnarray}
where $\kappa=(p_\mathrm{c}d/\hbar)^2$.
If we take such cutoff $p_\mathrm{c}$ that the filled valence band has two electrons per unit cell, as
in Section~\ref{sec:rect}, then $\kappa=4\pi/3\sqrt3$, and the kinetic pressure nearly halves in absolute value,
while the anomalous pressure drops by one quarter near the Dirac point. The next-order correction is much less
significant though.

We also consider the isotropic quadratic correction coming from the next-to-nearest-neighbor hopping,
$\delta\epsilon_{\mathbf{p}\gamma}^{(2)}=\nu v_\mathrm{F} p^2d/\hbar$, where $\nu =3t_2/2t$, $t$ and $t_2$ are the
nearest-neighbor and the next-to-nearest-neighbor hopping integrals. The corresponding pressure corrections are:
\begin{eqnarray}
\delta\mathcal{P}_\mathrm{kin}^{(2)}=\frac{g\nu\kappa^{1/2}v_\mathrm{F}}{8\pi\hbar^2}
\left(\frac{p_\mathrm{F}^4}{p_\mathrm{c}^4}-1\right)p_\mathrm{c}^3,\quad
\delta\mathcal{P}_\mathrm{anom}^{(2)}=\frac{g\nu\kappa^{1/2}v_\mathrm{F}}{4\pi\hbar^2}(s_\mu
p_\mathrm{F}^2+p_\mathrm{c}^2)p_\mathrm{c}.
\end{eqnarray}
If we take the upper bound $\nu=0.3$ \cite{CastroNeto}, then the correction approximately doubles the kinetic pressure
and increases the anomalous pressure by half.
\end{widetext}


\section*{References}

\begin{thebibliography}{99}

\bibitem{CastroNeto}
A.H. Castro Neto, F. Guinea, N.M.R. Peres, K.S. Novoselov, A.K. Geim, Rev. Mod. Phys. 81 (2009) 109--162.

\bibitem{Armitage}
N.P. Armitage, E.J. Mele, A. Vishwanath, Rev. Mod. Phys. 90 (2018) 015001.

\bibitem{Wehling}
T.O. Wehling, A.M. Black-Schaffer, A.V. Balatsky, Adv. Phys. 63 (2014) 1--75.

\bibitem{Marc}
G. Marc, W.G. McMillan, Adv. Chem. Phys. 58 (1985) 209--361.

\bibitem{Nielsen}
O.H. Nielsen, R.M. Martin, Phys. Rev. B 32 (1985) 3780--3791.

\bibitem{Maranganti}
R. Maranganti, P. Sharma, Proc. R. Soc. A 466 (2010) 2097--2116.

\bibitem{MartinPendas}
A. Mart\'{\i}n Pend\'as, J. Chem. Phys. 117 (2002) 965--979.

\bibitem{Fock}
V. Fock, Z. Phys. 63 (1930) 855--858.

\bibitem{Lowdin}
P.-O. L\"{o}wdin, J. Mol. Spectrosc. 3 (1959) 46--66.

\bibitem{Ziesche}
P. Ziesche, D. Lehmann, phys. stat. sol. (b) 139 (1987) 467--483.

\bibitem{Godfrey}
M.J. Godfrey,  Phys. Rev. B 37 (1988) 10176--10183.

\bibitem{Bader}
R.F.W. Bader, M.A. Austen, J. Chem. Phys. 107 (1997) 4271--4285.

\bibitem{Abad}
J. Abad, J.G. Esteve,  Phys. Rev. A 44 (1991) 4728--4729.

\bibitem{Bobrov}
V.B. Bobrov, S.A. Trigger, G.J.F. van Heijst, P.P.J.M. Schram, Phys. Rev. E 82 (2010) 010102(R).

\bibitem{Srebrenik}
S. Srebrenik, R.F.W. Bader, T.T. Nguyen-Dang, J. Chem. Phys. 68 (1978) 3667--3679.

\bibitem{Argyres}
P.N. Argyres, Int. J. Quant. Chem. 1S (1967) 669--675.

\bibitem{Esteve}
J.G. Esteve, F. Falceto, C. Garc\'{i}a Canal, Phys. Lett. A 374 (2010) 819--822.

\bibitem{Konstantinou}
G. Konstantinou, K. Kyriakou, K. Moulopoulos, Int. J. Eng. Innov. Res. 5 (2016) 248--252.

\bibitem{Cottrell}
T.L. Cottrell, S. Paterson, Philos. Mag. 42:327 (1951) 391--395.

\bibitem{Fernandez}
F.M. Fernandez, E.A. Castro, Int. J. Quantum Chem. 21 (1982) 741--751.

\bibitem{Stokes}
J.D. Stokes, H.P. Dahal, A.V. Balatsky, K.S. Bedell, Philos. Mag. Lett. 93 (2013) 672--679.

\bibitem{Sokolik}
A.A. Sokolik, A.D. Zabolotskiy, Yu.E. Lozovik, Phys. Rev. B 93 (2016) 195406.

\bibitem{Brey}
L. Brey, H.A. Fertig, Phys. Rev. B 73 (2006) 235411.

\bibitem{Berry}
M.V. Berry, R.J. Mondragon, Proc. R. Soc. Lond. A 412 (1987) 53--74.

\bibitem{McCann}
E. McCann, V.I. Fal'ko, J. Phys. Condens. Matter 16 (2004) 2371--2379.

\bibitem{Akhmerov}
A.R. Akhmerov, C.W.J. Beenakker, Phys. Rev. B 77 (2008) 085423.

\bibitem{Volkov}
V.A. Volkov, V.V. Enaldiev, JETP 122 (2016) 608--620.

\bibitem{Hashimoto}
K. Hashimoto, T. Kimura, X. Wu, Prog. Theor. Exp. Phys. 2017 (2017) 053I01.

\bibitem{Lin}
C.L. Lin, C.R. Ord\'{o}\~{n}ez, J. Stat. Mech. 2017 (2017) 043109.

\bibitem{Pereira}
V.M. Pereira, J. Nilsson, A.H. Castro Neto, Phys. Rev. Lett. 99 (2007) 166802.

\bibitem{Lozovik}
Yu.E. Lozovik, A.A. Sokolik, A.D. Zabolotskiy, Phys. Rev. B 91 (2015) 075416.

\bibitem{Peres}
N.M.R. Peres, F. Guinea, and A.H. Castro Neto, Phys. Rev. B 72 (2005) 174406.

\bibitem{Barlas}
Y. Barlas, T. Pereg-Barnea, M. Polini, R. Asgari, and A.H.MacDonald, Phys. Rev. Lett. 98 (2007) 236601.

\bibitem{Henry}
D. Henry, Perturbation of the Boundary in Boundary-Value Problems of Partial Differential Equations, Cambridge Univ.
Press, Cambridge, 2005, pp. 80--82.

\bibitem{Nakata}
Y. Nakata, Y. Ito, Y. Nakamura, and R. Shindou, https://arxiv.org/abs/1903.07052v1.

\end{thebibliography}
\end{document}